\documentclass[11pt]{article}
\usepackage{amsfonts}
\usepackage{graphicx}
\usepackage{authblk}
\usepackage{latexsym,amsmath}
\usepackage{amssymb,array}
\usepackage{enumerate}
\makeatletter \oddsidemargin 0in \evensidemargin 0in \textwidth 16cm
\DeclareMathOperator*{\ve}{\vee}
\begin{document}
\numberwithin{equation}{section}
\newtheorem{d1}{Definition}[section]
\newtheorem{thm}{Theorem}[section]
\title{\bf On Component Redundancy Versus System Redundancy for a $k$-out-of-$n$ System}
\author[a]{Mithu Rani Kuiti}
\author[b]{Nil Kamal Hazra}
\author[b]{Maxim Finkelstein\footnote{Corresponding author, email: FinkelM@ufs.ac.za}}
\affil[a]{Department of Operations Management, Indian Institute of Management Calcutta (IIMC), Diamond Harbour Road, Kolkata-700104, India}
\affil[b]{Department of Mathematical Statistics and Actuarial Science, University of the Free State, 339 Bloemfontein 9300, South Africa}
\date{September, 2017} 
\maketitle
\begin{abstract}
Precedence order is a natural type of comparison for random variables in numerous engineering applications (e.g., for the stress-strength modeling). In this note, we show that, for a $k$-out-of-$n$ system, redundancy at the component level is superior to that at the system level with respect to the stochastic precedence order. Cases of active and cold redundancy are considered. Similar results for other stochastic orders were intensively discussed in the literature. 
\end{abstract}
\section{Introduction and Preliminaries}\label{s1}
Allocation of redundant components into a system is one of the efficient methods to enhance reliability of engineering systems. The main question is: how and where to allocate the redundant components into the system so that the resultant system will be the optimum one in some suitable stochastic sense? The most popular types of redundancy in applications are active redundancy and cold redundancy. In case of active redundancy, the original component and the redundant components operate simultaneously (in parallel) at a full load. As a result, the system lifetime is the maximum of the lifetimes of the original component and the redundant components. On the other hand, cold redundancy describes a situation when a redundant component starts operating only when the corresponding original component fails. Thus, the system lifetime is just the sum of the lifetimes of the original component and the redundant components.
\\\hspace*{0.3 in}A classical result of Barlow and Proschan~\cite{bp1}, states that active redundancy at the component level is superior to that at the system level with respect to the usual stochastic order, which is the most popular stochastic order in applications. Recall that the usual stochastic order for two random variables holds when the survival function describing one random variable is larger (or equal) than that for the other one at each point of support.  Later on, this result has been generalized and also extended in many different directions including other types of stochastic orderings (see Barlow and Proschan~\cite{bp1}, Boland and El-Neweihi~\cite{be}, Gupta and Nanda~\cite{gn}, Misra et al.~\cite{mdg}, Brito et al.~\cite{bzv}, Nanda and Hazra~\cite{nh}, Hazra and Nanda~\cite{hn}, Gupta and Kumar~\cite{gk}, Zhao et al.~\cite{zzl}, and the references there in). However, due to mathematical complexity, the more general systems (e.g., $k$-out-of-$n$ system, coherent system) with non-iid components have not been studied sufficiently. Although there exists some results for general systems, all of them are either for systems with iid components or systems with matching spares (cf. Misra et al.~\cite{mdg}). See also some important generalizations in recent papers  by Da and Ding~\cite{dd}, and Zhang et al.~\cite{zad}. Most of these results are obtained with respect to active redundancy, whereas in the current paper the case of the cold redundancy is also considered.
\\\hspace*{0.3 in} It should be noted that stochastic ordering are very useful tool to compare the lifetimes of two systems. Many different types of stochastic orders (for example, usual stochastic order, hazard rate order, reversed hazard rate order, etc.) have been developed in the literature in order to handle various comparison problems (see Shaked and Shanthikumar~\cite{shak} for encyclopaedic information on this topic). Each stochastic order has its own merit and more suitable applications. However, the stochastic precedence order was not considered before in the literature with respect to the optimal redundancy allocation and our paper fills this gap to some extent considering this order for the $k$-out-of-$n$ system with non-identical components. From a general point of view, we feel that this order was not sufficiently studied in the literature so far and needs more attention in stochastic community. One of the possible reasons for that is that comparison with respect to precedence order requires usually different stochastic technique than that used for other orders. This will be demonstrated by the proofs of our main results. For completeness, we first state its definition and then briefly discuss the usefulness of this order. For more details, we refer the reader to Boland et al.~\cite{bsc}, and Finkelstein~\cite{f}. 

  \begin{d1}
   Let $U$ and $V$ be two nonnegative random variables. Then, $U$ is said to be greater than $V$ in stochastic precedence (sp) order, denoted as $U\geq_{sp}V$, if
   \begin{eqnarray*}
   P(U>V)> P(V>U).
   \end{eqnarray*}
    We write $U=_{sp}V$ if $ P(U>V)= P(V>U).$
   \end{d1}

 Thus, the definition of this order is very simple and meaningful. In essence, it also says that  $P(U>V)>0.5$. It is relevant to numerous engineering applications when e.g., stress-strength (Finkelstein~\cite{f})  or peak over the threshold problems are considered. In this type of problems, the precedent order is definitely  the most natural one as it directly describes the probability of interest (distinct from other popular stochastic orders). It can be easily shown that for the specific case of independent $U$ and $V$, the precedence order follows from the usual stochastic order. It is weaker and more flexible and can describe random variables with crossing reliability functions. On the other hand, in this paper, we are dealing with dependent $U$ and $V$, as the lifetimes of systems with different redundancy allocations obviously dependent. 
\\\hspace*{0.3 in}  To summarize: our goal in this paper is to study the optimal redundancy allocation for a $k$-out-of-$n$ system of non-identical components with respect to the stochastic precedence order. In what follows, we consider both active and cold redundancy.  Let us describe now formally the system and relevant notation. 
\\\hspace*{0.3 in}Consider a $k$-out-of-$n$ system with lifetime $\tau_{k:n}$ formed by $n$ components ${\mbox{\boldmath $Z$}}=(Z_1,Z_2,\dots,Z_n)$.
 Further, let ${\mbox{\boldmath $z$}}(t)\in \{0,1\}^n$ be the state vector of ${\mbox{\boldmath $Z$}}$, where $z_i(t)=1$ if the $i$th component is operating and $z_i(t)=0$ if it is not operating at time $t$.
 Without any loss of generality, we write ${\mbox{\boldmath $z$}}$ in place of ${\mbox{\boldmath $z$}}(t)$, for notational simplicity, when there is no ambiguity.  
 Then, the state of $\tau_{k:n}{({\mbox{\boldmath $Z$}})}$ at time $t$, is defined as
\begin{align*}
\phi_{\tau_{k:n}({\mbox{\boldmath $z$}})}=
  \left \{
      \begin{array}{ll}
            1, & \text{if the system is functioning}\\
            0, & \text{ if the system is not functioning}.
       \end{array}
 \right.
\end{align*}
  \begin{figure}
\centering
\includegraphics[width=16 cm, height=6 cm]{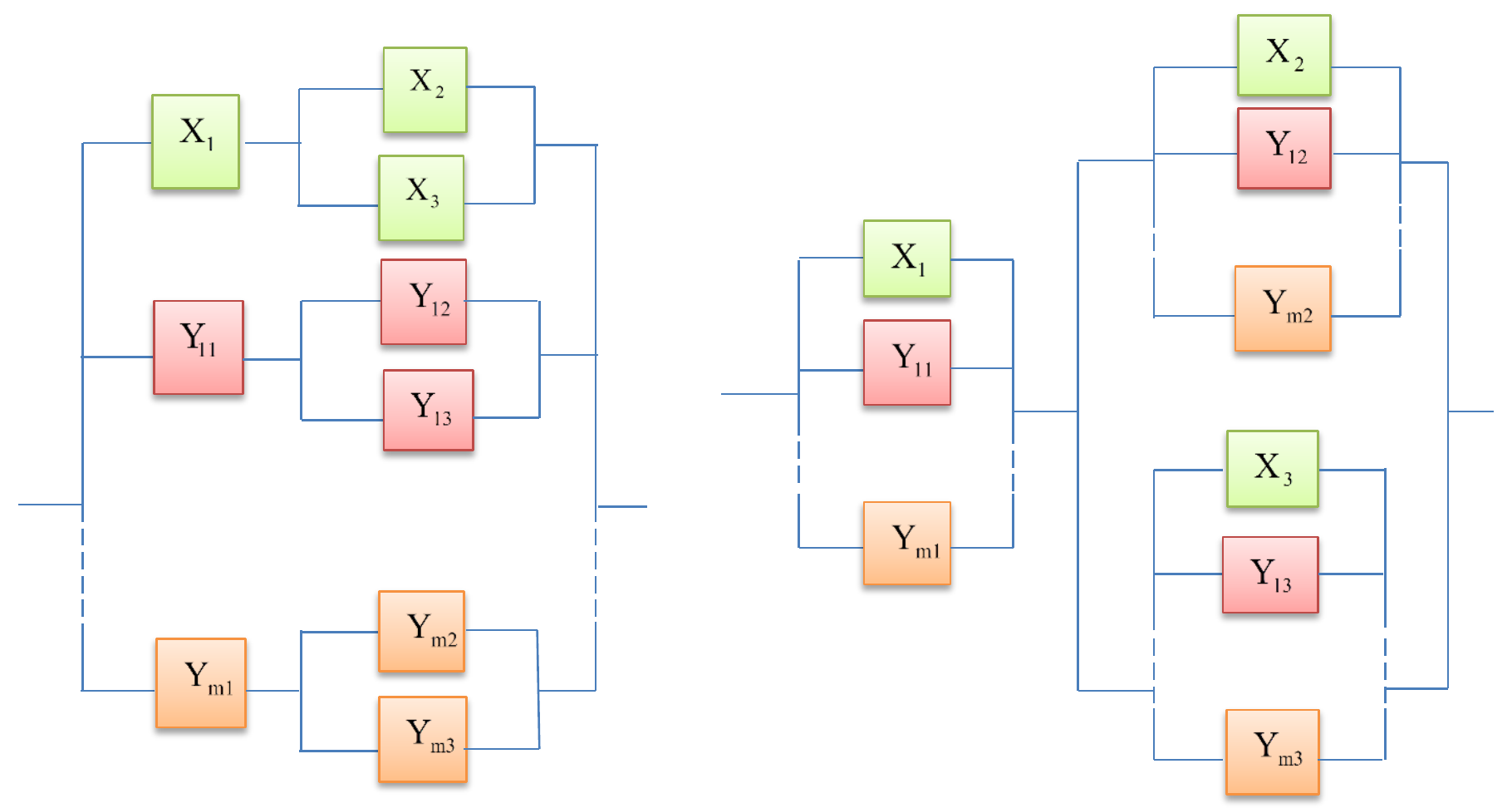} 
\caption{System redundancy and component redundancy, respectively} \label{f}
\end{figure}
Let ${\mbox{\boldmath $X$}}=(X_1,X_2,\dots,X_n)$ be a vector of random variables representing the lifetimes of $n$ components. Further, let $\{{\mbox{\boldmath $Y$}}_1,{\mbox{\boldmath $Y$}}_2,\dots,{\mbox{\boldmath $Y$}}_m\}$ be $m$ different sets of random variables representing the lifetimes of $mn$ number of redundancies, where ${\mbox{\boldmath $Y$}}_i=(Y_{i1},Y_{i2},\dots,Y_{in})$, for $i=1,2,\dots,m$. Assume that all $X_j$'s and $Y_{ij}$'s are independent, for all $i=1,2,\dots,m$ and $j=1,2,\dots,n$. We write $\tau_{k:n}\left({\mbox{\boldmath $X \vee Y_1\vee  Y_2\vee  \dots \vee  Y_m$}}\right)$ to denote the lifetime of a $k$-out-of-$n$ system with active redundancy at the component level, where $({\mbox{\boldmath $X \vee Y_1\vee  Y_2\vee  \dots \vee  Y_m$}})$ represents a $n$-tuple vector $(\max\{X_1,Y_{11},Y_{21},Y_{m1}\},\max\{X_2,Y_{12},Y_{22},Y_{m2}\},\dots,\max\{X_n,Y_{1n},Y_{2n},Y_{mn}\})$. Further, we write
$\tau_{k:n}({\mbox{\boldmath $X$}})\vee  \tau_{k:n}({\mbox{\boldmath $Y$}}_1)\vee  \tau_{k:n}({\mbox{\boldmath $Y$}}_2)\vee  \dots \vee \tau_{k:n}({\mbox{\boldmath $Y$}}_m)$ to represent the lifetime of a $k$-out-of-$n$ system with active redundancy at the system level, where the symbol $\vee$ stands for maximum.
Similarly, we write $\tau_{k:n}\left({\mbox{\boldmath $X + \sum_{i=1}^m Y_i$}}\right)$ to denote the lifetime of a $k$-out-of-$n$ system with cold redundancy at the component level, where $\left({\mbox{\boldmath $X + \sum_{i=1}^m Y_i$}}\right)$ represents a $n$-tuple vector $(X_1+\sum_{i=1}^m Y_{i1},X_2+\sum_{i=1}^m Y_{i2},\dots,X_n+\sum_{i=1}^m Y_{in})$. Further, we write
$\tau_{k:n}({\mbox{\boldmath $X$}})+\sum_{i=1}^m  \tau_{k:n}({\mbox{\boldmath $Y$}}_i)$ to represent the lifetime of a $k$-out-of-$n$ system with cold redundancy at the system level. An illustration of system redundancy versus component redundancy for a $3$-out-of-$3$ system is given in Figure~\ref{f}.

  \hspace*{0.3 in}The rest of the note is organized as follows. In Section~\ref{se2}, we discuss the main results of this paper. We show that for a $k$-out-of-$n$ system, redundancy at the component level is better than that at the system level with respect to the stochastic precedence order. The short concluding remarks are given in Section~\ref{se3}.
   
\section{Main Results}\label{se2}
Below we show that for a $k$-out-of-$n$ system, active redundancy at the component level is better than that at the system level with respect to the stochastic precedence order.
\begin{thm}\label{th.1}
Let ${\mbox{\boldmath $X$}}$ and $\{{\mbox{\boldmath $Y$}}_1,{\mbox{\boldmath $Y$}}_2,\dots,{\mbox{\boldmath $Y$}}_m\}$ be same as discussed in Section~\ref{s1}. Then, 
\begin{eqnarray*}
\tau_{1:n}\left({\mbox{\boldmath $X \vee Y_1\vee  Y_2\vee  \dots \vee  Y_m$}}\right)=_{sp}\tau_{1:n}({\mbox{\boldmath $X$}})\vee  \tau_{1:n}({\mbox{\boldmath $Y$}}_1)\vee  \tau_{1:n}({\mbox{\boldmath $Y$}}_2)\vee  \dots \vee \tau_{1:n}({\mbox{\boldmath $Y$}}_m),
\end{eqnarray*}
and, for $k=2,3,\dots, n $, 
\begin{eqnarray*}
\tau_{k:n}\left({\mbox{\boldmath $X \vee Y_1\vee  Y_2\vee  \dots \vee  Y_m$}}\right)\geq_{sp}\tau_{k:n}({\mbox{\boldmath $X$}})\vee  \tau_{k:n}({\mbox{\boldmath $Y$}}_1)\vee  \tau_{k:n}({\mbox{\boldmath $Y$}}_2)\vee  \dots \vee \tau_{k:n}({\mbox{\boldmath $Y$}}_m). 
\end{eqnarray*}
\end{thm}
{\bf Proof:} Let ${\mbox{\boldmath $x$}}=\left(x_1, x_2,\dots, x_n\right)$ and ${\mbox{\boldmath $y$}}_i=\left(y_{i1},y_{i2},\dots,y_{in}\right)$, $i=1,2,\dots, m$, be the state vectors of  ${\mbox{\boldmath $X$}}$ and ${\mbox{\boldmath $Y$}}_i$, respectively. Then, $x_j, y_{ij}\in \{0,1\}$, for all $i=1,2,\dots, m $ and $j=1,2,\dots,n$. Now, we first show that 
\begin{eqnarray}
\phi_{\tau_{k:n}({\mbox{\boldmath $x$}})\vee  \tau_{k:n}({\mbox{\boldmath $y$}}_1)\vee  \tau_{k:n}({\mbox{\boldmath $y$}}_2)\vee  \dots \vee \tau_{k:n}({\mbox{\boldmath $y$}}_m)}> \phi_{\tau_{k:n}\left({\mbox{\boldmath $x \vee y_1\vee  y_2\vee  \dots \vee  y_m$}}\right)}\label{eq1}
\end{eqnarray}
is never possible. Note that, $(\ref{eq1})$ holds if, and only if,
\begin{equation}\label{eq2}
\left\{ \begin{array}{l l}
\phi_{\tau_{k:n}({\mbox{\boldmath $x$}})\vee  \tau_{k:n}({\mbox{\boldmath $y$}}_1)\vee  \tau_{k:n}({\mbox{\boldmath $y$}}_2)\vee  \dots \vee \tau_{k:n}({\mbox{\boldmath $y$}}_m)}=1\\
\phi_{\tau_{k:n}\left({\mbox{\boldmath $x \vee y_1\vee  y_2\vee  \dots \vee  y_m$}}\right)}=0,
\end{array}
\right.
\end{equation}
or equivalently, one of the following five cases holds.
\\ Case I: 
Let $\phi_{\tau_{k:n}({\mbox{\boldmath $x$}})}=1$ and $\phi_{\tau_{k:n}({\mbox{\boldmath $y$}}_i)}=0$, $\text{for all } i=1,2,\dots,m$. Then, (\ref{eq2}) can equivalently be written as
\begin{eqnarray*}
&&x_1+x_2+\dots+x_n\geq k \\
&&y_{i1}+y_{i2}+\dots+y_{in}\leq {k-1}, \text{ for all } i=1,2,\dots,m \\
&&
\ve\limits_{i=1}^m\left(x_1, y_{i1}\right)+\ve\limits_{i=1}^m\left(x_2, y_{i2}\right)+\dots+\ve\limits_{i=1}^m\left(x_n, y_{in}\right)\leq {k-1}.
\end{eqnarray*}
Case II: Let $\phi_{\tau_{k:n}({\mbox{\boldmath $x$}})}=1$ and $\phi_{\tau_{k:n}({\mbox{\boldmath $y$}}_i)}=1$, $\text{for all } i=1,2,\dots,m$. Then, (\ref{eq2}) can equivalently be written as
\begin{eqnarray*}
&&x_1+x_2+\dots+x_n\geq k \\
&&y_{i1}+y_{i2}+\dots+y_{in}\geq k, \text{ for all } i=1,2,\dots,m\\
&&\ve\limits_{i=1}^m\left(x_1, y_{i1}\right)+\ve\limits_{i=1}^m\left(x_2, y_{i2}\right)+\dots+\ve\limits_{i=1}^m\left(x_n, y_{in}\right)\leq {k-1}.
\end{eqnarray*}
Case III: For each $r=1,2,\dots,m-1$, let $\phi_{\tau_{k:n}({\mbox{\boldmath $x$}})}=1$, and $\phi_{\tau_{k:n}({\mbox{\boldmath $y$}}_i)}=1$, $\text{for all } i=i_1,i_2,\dots,i_r$, and $\phi_{\tau_{k:n}({\mbox{\boldmath $y$}}_i)}=0$, $\text{for all } i=i_{r+1},i_{r+2},\dots,i_m$, where $\{i_1,i_2,\dots,i_r,i_{r+1},\dots,i_m \}\equiv\{1,2,\dots,m\}$. Then, (\ref{eq2}) can equivalently be written as
\begin{eqnarray*}
&&x_1+x_2+\dots+x_n\geq k \\
&&y_{i1}+y_{i2}+\dots+y_{in}\geq k, \text{ for all } i=i_1,i_2,\dots,i_r\\
&&y_{i1}+y_{i2}+\dots+y_{in}\leq k-1, \text{ for all } i=i_{r+1},i_{r+2},\dots,i_m\\
&&\ve\limits_{i=1}^m\left(x_1, y_{i1}\right)+\ve\limits_{i=1}^m\left(x_2, y_{i2}\right)+\dots+\ve\limits_{i=1}^m\left(x_n, y_{in}\right)\leq {k-1}.
\end{eqnarray*} 
Case IV: Let $\phi_{\tau_{k:n}({\mbox{\boldmath $x$}})}=0$ and $\phi_{\tau_{k:n}({\mbox{\boldmath $y$}}_i)}=1$, $\text{for all } i=1,2,\dots,m$. Then, (\ref{eq2}) can equivalently be written as
\begin{eqnarray*}
&&x_1+x_2+\dots+x_n\leq k-1 \\
&&y_{i1}+y_{i2}+\dots+y_{in}\geq k, \text{ for all } i=1,2,\dots,m\\
&&\ve\limits_{i=1}^m\left(x_1, y_{i1}\right)+\ve\limits_{i=1}^m\left(x_2, y_{i2}\right)+\dots+\ve\limits_{i=1}^m\left(x_n, y_{in}\right)\leq {k-1}.
\end{eqnarray*} 
Case V: For each $r=1,2,\dots,m-1$, let $\phi_{\tau_{k:n}({\mbox{\boldmath $x$}})}=0$, and $\phi_{\tau_{k:n}({\mbox{\boldmath $y$}}_i)}=1$, $\text{for all } i=i_1,i_2,\dots,i_r$, and $\phi_{\tau_{k:n}({\mbox{\boldmath $y$}}_i)}=0$, $\text{for all } i=i_{r+1},i_{r+2},\dots,i_m$, where $\{i_1,i_2,\dots,i_r,i_{r+1},\dots,i_m \}\equiv\{1,2,\dots,m\}$. Then, (\ref{eq2}) can equivalently be written as
\begin{eqnarray*}
&&x_1+x_2+\dots+x_n\leq k-1 \\
&&y_{i1}+y_{i2}+\dots+y_{in}\geq k, \text{ for all } i=i_1,i_2,\dots,i_r\\
&&y_{i1}+y_{i2}+\dots+y_{in}\leq k-1, \text{ for all } i=i_{r+1},i_{r+2},\dots,i_m\\
&&\ve\limits_{i=1}^m\left(x_1, y_{i1}\right)+\ve\limits_{i=1}^m\left(x_2, y_{i2}\right)+\dots+\ve\limits_{i=1}^m\left(x_n, y_{in}\right)\leq {k-1}.
\end{eqnarray*}  
It can be verified that no system of inequalities given in above five cases has any solution. Thus,
\begin{eqnarray}
P\left[\tau_{k:n}({\mbox{\boldmath $X$}})\vee  \tau_{k:n}({\mbox{\boldmath $Y$}}_1)\vee  \tau_{k:n}({\mbox{\boldmath $Y$}}_2)\vee  \dots \vee \tau_{k:n}({\mbox{\boldmath $Y$}}_m)
\right.\nonumber
\\\left.>\tau_{k:n}\left({\mbox{\boldmath $X \vee Y_1\vee  Y_2\vee  \dots \vee  Y_m$}}\right)\right]=0.\label{eq3}
\end{eqnarray}
Now, we will verify whether 
$$P\left[\tau_{k:n}\left({\mbox{\boldmath $X \vee Y_1\vee  Y_2\vee  \dots \vee  Y_m$}}\right)
> \tau_{k:n}({\mbox{\boldmath $X$}})\vee  \tau_{k:n}({\mbox{\boldmath $Y$}}_1)\vee  \tau_{k:n}({\mbox{\boldmath $Y$}}_2)\vee  \dots \vee \tau_{k:n}({\mbox{\boldmath $Y$}}_m) \right]$$
is also zero or not. Below we show that this probability is not always zero. Note that, 
\begin{eqnarray*}\label{eq4}
\phi_{\tau_{k:n}\left({\mbox{\boldmath $x \vee y_1\vee  y_2\vee  \dots \vee  y_m$}}\right)}>\phi_{\tau_{k:n}({\mbox{\boldmath $x$}})\vee  \tau_{k:n}({\mbox{\boldmath $y$}}_1)\vee  \tau_{k:n}({\mbox{\boldmath $y$}}_2)\vee  \dots \vee \tau_{k:n}({\mbox{\boldmath $y$}}_m)}
\end{eqnarray*} 
holds if, and only if,
\begin{eqnarray*}
&&\phi_{\tau_{k:n}({\mbox{\boldmath $x$}})\vee  \tau_{k:n}({\mbox{\boldmath $y$}}_1)\vee  \tau_{k:n}({\mbox{\boldmath $y$}}_2)\vee  \dots \vee \tau_{k:n}({\mbox{\boldmath $y$}}_m)}=0
\\&&\phi_{\tau_{k:n}\left({\mbox{\boldmath $x \vee y_1\vee  y_2\vee  \dots \vee  y_m$}}\right)}=1.
\end{eqnarray*}
This is equivalent to the fact that
\begin{eqnarray*}
&&\phi_{\tau_{k:n}}({\mbox{\boldmath $x$}})=\phi_{\tau_{k:n}}({\mbox{\boldmath $y$}}_1)=\phi_{\tau_{k:n}}({\mbox{\boldmath $y$}}_2)\dots = \phi_{\tau_{k:n}}({\mbox{\boldmath $y$}}_m)=0\\&&
\phi_{\tau_{k:n}\left({\mbox{\boldmath $x \vee y_1\vee  y_2\vee  \dots \vee  y_m$}}\right)}=1,
\end{eqnarray*}
or equivalently, the following system of inequalities is satisfied.
\begin{eqnarray*}
&&x_1+x_2+\dots+x_n\leq {k-1} \\
&&y_{i1}+y_{i2}+\dots+y_{in}\leq {k-1}, \text { for all } i=1,2,\dots, m
\\&&\ve\limits_{i=1}^m\left(x_1, y_{i1}\right)+\ve\limits_{i=1}^m\left(x_2, y_{i2}\right)+\dots+\ve\limits_{i=1}^m\left(x_n, y_{in}\right)\geq k.
\end{eqnarray*}
It is to be noted that, the above system of inequalities has at least one solution except for $k=1$.
Thus,
\begin{eqnarray}\label{eq5}
P\left[\tau_{1:n}\left({\mbox{\boldmath $X \vee Y_1\vee  Y_2\vee  \dots \vee  Y_m$}}\right)
> \tau_{1:n}({\mbox{\boldmath $X$}})\vee  \tau_{1:n}({\mbox{\boldmath $Y$}}_1)\right.\nonumber
\\\left.\vee  \tau_{1:n}({\mbox{\boldmath $Y$}}_2)\vee  \dots \vee \tau_{1:n}({\mbox{\boldmath $Y$}}_m) \right]=0,
\end{eqnarray}
and for $k=2,3,\dots,n,$
\begin{eqnarray}\label{eq6}
P\left[\tau_{k:n}\left({\mbox{\boldmath $X \vee Y_1\vee  Y_2\vee  \dots \vee  Y_m$}}\right)
> \tau_{k:n}({\mbox{\boldmath $X$}})\vee  \tau_{k:n}({\mbox{\boldmath $Y$}}_1)\right.\nonumber
\\\left.\vee  \tau_{k:n}({\mbox{\boldmath $Y$}}_2)\vee  \dots \vee \tau_{k:n}({\mbox{\boldmath $Y$}}_m) \right]>0.
\end{eqnarray}
Therefore, on using (\ref{eq3}), (\ref{eq5}) and (\ref{eq6}), we have,
\begin{eqnarray*}
P\left[\tau_{k:n}\left({\mbox{\boldmath $X \vee Y_1\vee  Y_2\vee  \dots \vee  Y_m$}}\right)
> \tau_{k:n}({\mbox{\boldmath $X$}})\vee  \tau_{k:n}({\mbox{\boldmath $Y$}}_1)\vee  \tau_{k:n}({\mbox{\boldmath $Y$}}_2)\vee  \dots \vee \tau_{k:n}({\mbox{\boldmath $Y$}}_m) \right]
\\\geq 
P\left[\tau_{k:n}({\mbox{\boldmath $X$}})\vee  \tau_{k:n}({\mbox{\boldmath $Y$}}_1)\vee  \tau_{k:n}({\mbox{\boldmath $Y$}}_2)\vee  \dots \vee \tau_{k:n}({\mbox{\boldmath $Y$}}_m) >\tau_{k:n}\left({\mbox{\boldmath $X \vee Y_1\vee  Y_2\vee  \dots \vee  Y_m$}}\right)
\right],
\end{eqnarray*}
where the equality holds for $k=1$. Hence, the result follows. $\hfill\Box$
\\\hspace*{0.3 in}In the next theorem we consider cold redundancies in place of active redundancies. We show that the same result, as in the above theorem, also holds here.
\begin{thm}\label{th.2}
Let ${\mbox{\boldmath $X$}}$ and $\{{\mbox{\boldmath $Y$}}_1,{\mbox{\boldmath $Y$}}_2,\dots,{\mbox{\boldmath $Y$}}_m\}$ be same as discussed in Section~\ref{s1}. Then,
\begin{eqnarray*}
\tau_{1:n}\left({\mbox{\boldmath $X + \sum_{i=1}^m Y_i$}}\right)=_{sp}\tau_{1:n}({\mbox{\boldmath $X$}})+\sum_{i=1}^m  \tau_{1:n}({\mbox{\boldmath $Y$}}_i),
\end{eqnarray*}
and for $k=2,3,\dots,n,$
\begin{eqnarray*}
\tau_{k:n}\left({\mbox{\boldmath $X + \sum_{i=1}^m Y_i$}}\right)\geq_{sp}\tau_{k:n}({\mbox{\boldmath $X$}})+\sum_{i=1}^m  \tau_{k:n}({\mbox{\boldmath $Y$}}_i).
\end{eqnarray*}
\end{thm}
{\bf Proof:} Let ${\mbox{\boldmath $x$}}=\left(x_1, x_2,\dots, x_n\right)$ and ${\mbox{\boldmath $y$}}_i=\left(y_{i1},y_{i2},\dots,y_{in}\right)$, $i=1,2,\dots, m$, be the state vectors of  ${\mbox{\boldmath $X$}}$ and ${\mbox{\boldmath $Y$}}_i$, respectively. Then, $x_j, y_{ij}\in \{0,1\}$, and $(x_j,y_{ij})\neq (1,1)$ and $(y_{ij},y_{lj})\neq(1,1)$, for all $i,l=1,2,\dots, m $ and $j=1,2,\dots,n$, and $i\neq l$. Now, we first show that  
\begin{eqnarray}\label{eq7}
\phi_{\tau_{k:n}({\mbox{\boldmath $x$}})+\sum_{i=1}^m  \tau_{k:n}({\mbox{\boldmath $y$}}_i)}>\phi_{\tau_{k:n}\left({\mbox{\boldmath $x + \sum_{i=1}^m y_i$}}\right)}
\end{eqnarray}
is never possible. Note that, (\ref{eq7}) holds if, and only if,
\begin{equation}\label{eq8}
\left\{ \begin{array}{l l}
\phi_{\tau_{k:n}({\mbox{\boldmath $x$}})+\sum_{i=1}^m  \tau_{k:n}({\mbox{\boldmath $y$}}_i)}=1\\
\phi_{\tau_{k:n}\left({\mbox{\boldmath $x + \sum_{i=1}^m y_i$}}\right)}=0,
\end{array}
\right.
\end{equation}
or equivalently, one of the following two cases holds.
\\Case I: Let $\phi_{\tau_{k:n}({\mbox{\boldmath $x$}})}=1$ and $\phi_{\tau_{k:n}({\mbox{\boldmath $y$}}_i)}=0$, $\text{for all } i=1,2,\dots,m$. Then, (\ref{eq8}) can equivalently be written as
\begin{eqnarray*}
&&x_1+x_2+\dots+x_n\geq k
\\&&y_{i1}+y_{i2}+\dots+y_{in}\leq {k-1}\text{ for all } i=1,2,\dots m 
\\&&\left(x_1+\sum_{l=1}^{m}y_{l1}\right)+\left(x_2+\sum_{l=1}^{m}y_{l2}\right)+\dots+\left(x_n+\sum_{l=1}^{m}y_{ln}\right)\leq {k-1}.                
\end{eqnarray*}
Case II: For each $r=1,2,\dots,m$, let $\phi_{\tau_{k:n}({\mbox{\boldmath $x$}})}=0$, $\phi_{\tau_{k:n}({\mbox{\boldmath $y$}}_{r})}=1$, and $\phi_{\tau_{k:n}({\mbox{\boldmath $y$}}_i)}=0$,  $\text{for all } i=i_1,i_2,\dots,i_{m-1}$, where $\{i_1,i_2,\dots,\dots,i_{m-1} \}\equiv\{1,2,\dots,m\}\setminus r$. Then, (\ref{eq8}) can equivalently be written as
\begin{eqnarray*}
&&x_1+x_2+\dots+x_n\leq k-1
\\&&y_{r1}+y_{r2}+\dots+y_{rn}\geq {k} 
\\&&y_{i1}+y_{i2}+\dots+y_{in}\leq {k-1}, \text{ for all } i=i_1,i_2,\dots,i_{m-1}  
\\&&\left(x_1+\sum_{l=1}^{m}y_{l1}\right)+\left(x_2+\sum_{l=1}^{m}y_{l2}\right)+\dots+\left(x_n+\sum_{l=1}^{m}y_{ln}\right)\leq {k-1}.                
\end{eqnarray*}
 It can be verified that no system of inequalities given in above two cases has any solution. Thus,
\begin{eqnarray}\label{eq10}
P\left[\tau_{k:n}({\mbox{\boldmath $X$}})+\sum_{i=1}^m  \tau_{k:n}({\mbox{\boldmath $Y$}}_i)>\tau_{k:n}\left({\mbox{\boldmath $X + \sum_{i=1}^m Y_i$}}\right)\right]=0.
\end{eqnarray}
Now, we will verify whether $P\left[\tau_{k:n}\left({\mbox{\boldmath $X + \sum_{i=1}^m Y_i$}}\right)>\tau_{k:n}({\mbox{\boldmath $X$}})+\sum_{i=1}^m  \tau_{k:n}({\mbox{\boldmath $Y$}}_i) \right]$ is also zero or not. Below we show that this probability is not always zero. Note that, 
\begin{eqnarray*}
\phi_{\tau_{k:n}\left({\mbox{\boldmath $x + \sum_{i=1}^m y_i$}}\right)}>\phi_{\tau_{k:n}({\mbox{\boldmath $x$}})+\sum_{i=1}^m  \tau_{k:n}({\mbox{\boldmath $y$}}_i)}
\end{eqnarray*}
holds if, and only if,
\begin{eqnarray*}
&&\phi_{\tau_{k:n}({\mbox{\boldmath $x$}})+\sum_{i=1}^m  \tau_{k:n}({\mbox{\boldmath $y$}}_i)}=0
\\&&\phi_{\tau_{k:n}\left({\mbox{\boldmath $x + \sum_{i=1}^m y_i$}}\right)}=1.
\end{eqnarray*}
This is equivalent to the fact that
\begin{eqnarray*}
&&\phi_{\tau_{k:n}({\mbox{\boldmath $x$}})}=0
\\&& \phi_{ \tau_{k:n}({\mbox{\boldmath $y$}}_i)}=0, \text{ for all }i=1,2,\dots,m
\\&&\phi_{\tau_{k:n}\left({\mbox{\boldmath $x + \sum_{i=1}^m y_i$}}\right)}=1,
\end{eqnarray*}
or equivalently, the following system of inequalities is satisfied.
\begin{eqnarray*}
&&x_1+x_2+\dots+x_n\leq {k-1}
 \\&&y_{i1}+y_{i2}+\dots+y_{in}\leq {k-1}\text{ for all } i=1,2,\dots m
  \\&&\left(x_1+\sum_{l=1}^{m}y_{l1}\right)+\left(x_2+\sum_{l=1}^{m}y_{l2}\right)+\dots+\left(x_n+\sum_{l=1}^{m}y_{ln}\right)\geq k               
\end{eqnarray*}
It is to be noted that the above system of inequalities has at least one solution except for $k=1$.
Thus,
\begin{eqnarray}\label{eq18}
P\left[\tau_{1:n}\left({\mbox{\boldmath $X + \sum_{i=1}^m Y_i$}}\right)>\tau_{1:n}({\mbox{\boldmath $X$}})+\sum_{i=1}^m  \tau_{1:n}({\mbox{\boldmath $Y$}}_i) \right]=0,
\end{eqnarray} 
and for $k=2,3,\dots,n$, 
\begin{eqnarray}\label{eq9}
P\left[\tau_{k:n}\left({\mbox{\boldmath $X + \sum_{i=1}^m Y_i$}}\right)>\tau_{k:n}({\mbox{\boldmath $X$}})+\sum_{i=1}^m  \tau_{k:n}({\mbox{\boldmath $Y$}}_i) \right]>0.
\end{eqnarray}
Therefore, on using (\ref{eq10}), (\ref{eq18}) and (\ref{eq9}), we have,
\begin{eqnarray*}
&&P\left[\tau_{k:n}\left({\mbox{\boldmath $X + \sum_{i=1}^m Y_i$}}\right)>\tau_{k:n}({\mbox{\boldmath $X$}})+\sum_{i=1}^m  \tau_{k:n}({\mbox{\boldmath $Y$}}_i) \right]
\\&&\geq P\left[\tau_{k:n}({\mbox{\boldmath $X$}})+\sum_{i=1}^m  \tau_{k:n}({\mbox{\boldmath $Y$}}_i)>\tau_{k:n}\left({\mbox{\boldmath $X + \sum_{i=1}^m Y_i$}}\right)\right],
\end{eqnarray*}
where the equality holds for $k=1$. Hence, the result follows.$\hfill\Box$
\section{Concluding Remarks}\label{se3}
In this note, we study stochastic comparisons between the systems with redundancy on the component and  the system level. We show that, for a $k$-out-of-$n$ system, allocation of redundant components  at the component level is superior to that at the system level with respect to the stochastic precedence order.
As the precedence order is the most natural in numerous engineering applications, we believe that this result can help in structural decision making in various practical situations. In the future research,  we plan to generalize the obtained results to the case of general coherent systems, which is the major challenge.
\section*{Acknowledgements}
\hspace*{0.3 in}The first author sincerely acknowledges the financial support from the Indian Institute of Management Calcutta, India. The work of the second author is supported by Claude Leon Foundation, South Africa.


\end{document}